\documentclass[12pt]{article}

\usepackage{times}

\newenvironment{sciabstract}{%
\begin{quote} \bf}
{\end{quote}}

\newcounter{lastnote}
\newenvironment{scilastnote}{%
\setcounter{lastnote}{\value{enumiv}}%
\addtocounter{lastnote}{+1}%
\begin{list}%
{\arabic{lastnote}.}
{\setlength{\leftmargin}{.22in}}
{\setlength{\labelsep}{.5em}}}
{\end{list}}

\title{Spinning Holes in Semiconductors}

\author
{H.A. Fertig$^{\ast}$ \\
\\
\normalsize{Department of Physics and Astronomy, University of Kentucky,}\\
\normalsize{Lexington, KY 40506-0055 USA}\\
\\
\normalsize{$^\ast$E-mail:  fertig@pa.uky.edu.}
}

\date{}

\begin{document} 


\baselineskip24pt


\maketitle


\begin{sciabstract}
The electron spin is emerging as a new powerful tool in the electronics
and optics industries.  Many proposed applications involve the
creation of spin currents, which so far have proven to be difficult
to produce in semiconductor environments.   A new theoretical
analysis shows this might be achieved using holes rather than
electrons in semiconductors with significant spin-orbit coupling.
\end{sciabstract}

Perhaps the most prominent characteristic of the electron is the fact that
it carries electric charge.  Together with the rules of quantum mechanics,
the electric forces among electrons and nuclei
determine the chemical properties of atoms and molecules.  Manipulation
of electrons in semiconductors using 
electric forces, which couple to the
charge,
is the principle behind the revolution in the electronics
industry of the last few decades.

Another fundamental feature of electrons is their spin, a property in which
the charge apparently spins like a top, endowing the electron
with a magnetic dipole moment much like that of a bar magnet.  
Incorporating and exploiting this spin in microelectronic and optoelectronic
applications is the central idea of {\it spintronics}\cite{wolf}.
While a number of commercially successful applications of
this already exist (most prominently as memory
for computers), many proposed future applications await
the development of methods to produce and manipulate spin currents.
An important theoretical step in this direction by Shuichi Murakami,
Naoto Nagaosa, and Shou-Cheng Zhang is reported in this
issue of Science \cite{murakami}.  

Unlike charge, electron spin is specified by a direction through its
rotation axis.  If one tries to measure the direction of this spin -- say,
by passing the electrons through a magnetic field gradient -- one finds
that the spin will point either ``up'' or ``down''; the rules of quantum
mechanics forbid any other result upon measurement.  One
could thus imagine using the spin as a bit in a computer, with a down
spin state representing 0 and up representing 1.  Quantum mechanics
however allows much richer possibilities than this.  The electron
spin can be in a state that is not just up or down, but one that is 
a combination of the two.  The full range of possibilities may be 
represented by an arrow directed
toward any point on  a ``Bloch sphere'' \cite{nielsen}
(see Fig. 1).  It is
only upon measurement of the spin component along some
direction that quantum mechanics allows only two possible results.

This richness of possible states makes electron spin an ideal candidate
for a {\it qubit}, the basic component of the (as yet undeveloped)
quantum computer.   Quantum computers exploit
the quantum dynamics of spins to vastly improve the
speeds of tasks such as Fourier transformation and factorization
of large integers, which can often not be
performed by existing digital technology on
reasonable time scales.  Factorization in particular plays
a key role in cryptographic schemes, so 
government security agencies around the world have a
keen interest in quantum computers.

Materials that support spin currents can play a crucial role in the
practical development of quantum computers.   While there are
many proposals for systems that could support spins or
their analogs as qubits, one also needs practical
means to initialize the spin states as well as read them.
In semiconductor-based proposals for quantum computers,
such as quantum dots\cite{loss}, one can
use interactions between a spin-current carrying wire and
a qubit to read the qubit state, and ``spin-injection'' to
initialize it.  Moreover, qubits need to
interact in ways that do not dissipate the information stored
in their quantum states (as happens when an electron spin
is directly measured).  Spin currents have been demonstrated
to preserve their coherence over remarkably long distances
and times\cite{malajovich}, so materials capable of supporting
them could provide a medium through which the dots could
interact in a controllable manner.

One possible approach to creating spin currents is to exploit
spin-orbit
coupling, an effect in which the trajectory of an electron moving under
the influence of an electric field depends on its spin state.  For
example,
a recent proposal \cite{streda} suggests passing electrons through a
heterostructure engineered so that spin-orbit coupling might
be made relatively strong, generating a spin current perpendicular to
the electric current.  Murakami et al. demonstrate that
spin currents via spin-orbit coupling can be generated more simply
using holes rather than electrons, because relatively strong spin-orbit
coupling naturally exists for holes in many semiconducting systems
in which it is small or absent for electrons.  This idea offers several
practical
advantages.

First, many of the materials needed are commonly available and
can easily be processed.  Second,
because the direction of spin and 
electric currents are connected, the information
carried by the spin currents 
could in principle be translated into
normal electric currents.  This would facilitate the integration
of spintronics with traditional microelectronic
devices.  The use of common
semiconducting materials is a further benefit
in developing such integrated devices.  Finally, 
the polarization of the
current is fixed not by a magnetic field but
by the direction of the currents themselves, obviating
the need for magnets to fix the spin polarization.  This 
property may prove important in miniaturized
systems, where one may not wish to have magnetic fields
in every part of a given device.  

Once spin currents can be created and manipulated in this way,
quantum computers will arguably represent their most exciting
possible application.   Other applications
may emerge much sooner,  including spin
diodes and transitors\cite{flatte}, which could be at the
heart of high speed reprogrammable
logic circuit elements and 
non-volatile memory applications, 
electro-optic light modulators\cite{datta},
and circularly polarized light emitting diodes\cite{fiederling}.

\bibliography{scibib}

\bibliographystyle{Science}


\begin{scilastnote}
\item The author acknowledges the support of the NSF through
Materials Theory Grant No. DMR-0108451.
\end{scilastnote}


\clearpage

\noindent {\bf Figure.} An electron spin may be represented by an arrow,
and its quantum state specified by a point on a spherical surface towards which
the arrow points.  A measurement of the spin along
some direction (e.g., one of the coordinate axes) always results in
the spin being parallel or antiparallel to the measurement direction.
The quantum state determines the probability for each of these two results.
By allowing multiple spins to interact without directly measuring them,
the full range of possible states would be exploited by a quantum computer.

\end{document}